# SR-NCL: An Area-/Energy Efficient Resilient NCL Architecture based on Selective Redundancy

Hasnain A. Ziad
*Electrical and Computer Engineering*
*Florida Polytechnic University*
Lakeland, FL, USA
hziad7099@floridapoly.edu

Alexander C. Bodoh
*Electrical and Computer Engineering*
*Florida Polytechnic University*
Lakeland, FL, USA
abodoh8714@floridapoly.edu

Ashiq A. Sakib
*Electrical and Computer Engineering*
*Southern Illinois University Edwardsville*
Edwardsville, IL, USA
asakib@siue.ed

***Abstract*—Duplication-based redundancy schemes have proven to be effective in designing fully-resilient Quasi-delay Insensitive (QDI) asynchronous circuits. The complete resiliency, however, is accompanied by significant energy, latency, and area overhead. This paper presents a novel error-tolerant Null Convention Logic (NCL) architecture based on selective redundancy. Results demonstrate the efficacy of the proposed method in terms of area and energy utilization as compared to existing duplication-based NCL designs, targeting an image processing application.**

## I. Introduction

Digital integrated circuits (ICs) designed to withstand extreme environmental conditions are crucial for several applications of significant national interest, such as outer space exploration, deep-sea research, military surveillance, etc. However, reliability of such ICs is a major concern. Electronic components are becoming increasingly susceptible to transient errors caused by radiation, electromagnetic interferences (EMI), and/or unanticipated noise sources due to aggressive device and voltage scaling [1]. In a radiation-intensive environment, a transient current surge may occur when high-energy neutrons or alpha particles impact a sensitive circuit node. This event may subside fairly quickly or propagate, and may cause undesired switching in logic components, giving rise to a phenomenon called Single-Event Upset (SEU) [2]. SEUs pose greater threats in conventional clock-driven digital designs due to their increased susceptibility to process, voltage, and temperature (PVT) variations at scaled technology nodes. SEUs have the potential to alter the timing behavior of the clocked circuits, which can lead to desynchronization and eventually, circuit malfunction. The Quasi-delay Insensitive (QDI) asynchronous domain presents itself as a more suitable alternative for applications operating in harsh conditions, owing to its inherent robustness against PVT variations and insensitivity to timing discrepancies [3-4].

Although QDI circuits are robust, they are not completely SEU resistant. Numerous radiation-hardening schemes have been proposed over the years, of which duplication-based redundancy schemes have been found to be the most effective across all major QDI paradigms. However, complete resiliency comes at the cost of a significant area, energy, and performance overhead [5]. Our research is predicated on the premise that not all applications necessitate complete fault tolerance. Image and video processing systems serve as prudent examples that allow for a margin of error, as slight distortions in audio or visuals may go unnoticed due to perceptual limitations. This is the foundation of approximate computing, which we leverage in this research. Herein, we present a novel SEU-tolerant QDI Null Convention Logic (NCL) architecture based on selective redundancy, where only a segment of the overall circuit is duplicated. The objective is to enhance the area and energy efficiency of the circuit by deliberately reducing the SEU resilience to a level that is still acceptable for the target application, while preserving the NCL protocols and deadlock-free operation.

This rest of the paper is organized as follows. Section II briefly presents the original NCL framework and discusses some of the relevant error tolerant QDI designs. Section III details the proposed architecture and demonstrates the error-response of the design under different SEU scenarios, followed by the simulation results and comparison in Section IV. Section V concludes the paper.

## II. Background

### A. NCL overview

The NCL framework is depicted in Fig. 1 [6]. Each NCL pipeline stage consists of a set of input and output registers, a QDI combinational logic (*CL*) unit, and a completion detection (*CD*) unit. The *CL* unit implements the functionality, while the registration and *CD* units establish the control path for synchronization in the absence of a reference clock signal. NCL most commonly utilizes dual-rail logic, where each signal has a 2-bit encoding. A dual-rail signal, *D*, consists of

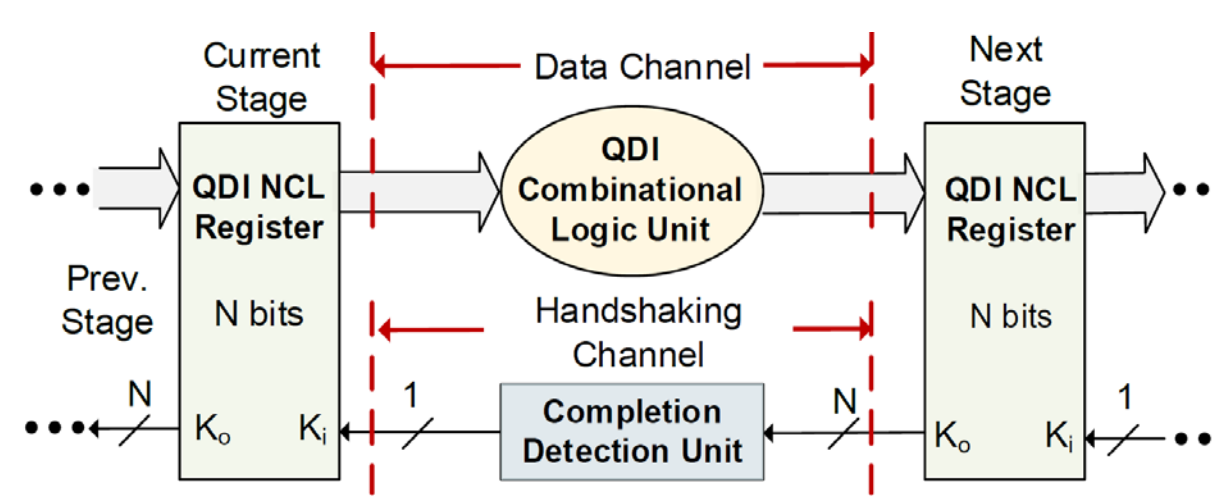

Fig. 1: NCL framework

This article has been accepted for publication in the Proceedings of the 2025 IEEE International Symposium on Circuits and Systems (ISCAS).

Citation: H. A. Ziad, A. C. Bodoh and A. A. Sakib, "SR-NCL: An Area-/Energy Efficient Resilient NCL Architecture based on Selective Redundancy," 2025 IEEE International Symposium on Circuits and Systems (ISCAS), London, United Kingdom, 2025, pp. 1-5, doi: 10.1109/ISCAS56072.2025.11043864.

two wires, $D^0$ and $D^1$, where ($D^1$, $D^0$) ∈ {0,1} and can have one of the three legal values from the set {NULL, DATA0, DATA1}. DATA0 ($D^1D^0=01$) and DATA1 ($D^1D^0=10$) are equivalent to Boolean logic '0' and '1', respectively. ($D^1D^0=00$) indicates a NULL state that acts as an intermediate state between two data fronts. ($D^1D^0=11$) is an illegal state. The registers, *CL* unit, and *CD* unit comprise of NCL threshold gates with state-holding (hysteresis) capability. The registration and completion units are utilized to implement a four-phased NCL handshaking protocol, that maintains an alternating sequence of NULL and DATA to differentiate between two valid data sets [6].

### B. Error Tolerant NCL Circuits

The error response of QDI circuits is distinct from that of their synchronous counterparts and therefore, the resiliency techniques that are applicable to synchronous designs are not readily applicable to QDI circuits [5]. Existing NCL resiliency schemes can be broadly classified into two categories: redundancy-based and non-redundancy-based. Redundancy-based schemes necessitate the replication of the original circuit and the implementation of additional logic for synchronization. These schemes offer complete resilience and have been applied to all major QDI paradigms, both at the gate and architectural level [7-12]. However, this resilience comes at the expense of increased energy, latency, and area overhead. The non-redundancy-based schemes demonstrate better performance in terms of area and energy utilization; however, they have other major limitations. For example, Kuang et al. proposed one such non-duplicating method in [13] for the design of a self-correcting NCL *CL* unit, extending upon the idea proposed in [14]. While the method can recover from errors that generate illegal data during the DATA phase, it is not effective during the NULL phase. Moreover, erroneous data could still propagate through the pipeline, necessitating additional mechanisms for discarding. Lodhi et al. addressed these limitations in [13] and proposed modifications in [15]. However, both [13] and [15] only focus on hardening the CL unit. Registration and *CD* units are presumed to be error-free, which is an oversimplification, as errors in registration and/or *CD* unit can cause circuit deadlock. In addition, both methods mandate the stalling of the pipeline during the re-computation phase of recovery, affecting the latency [5].

## III. Selective Redundancy based Error-Resilient NCL Circuit (SR-NCL) Design

Fig. 2 illustrates the proposed error tolerant NCL architecture. This method involves the duplication of a selected portion of the circuit, as opposed to previous modular redundancy methods, which involved the duplication of the entire circuit [10]. Consequently, the proposed architecture exhibits several unique distinctions from the existing Dual Modular Redundancy (DMR) based methods. The proposed method requires the following modifications:

- The NCL *CL* unit is partitioned, and a selective portion is duplicated, while the rest remains unaltered. The proposed architecture ensures that the duplicated portion remains entirely protected against SEUs, as detailed later.
- Each registration and *CD* circuitry is duplicated.
- In each pipeline stage, a single-level layer of *N* additional TH22 gates is added at the output of the *N*-bit input registers, similar to [10] and [12]. Each TH22 gate obtains its inputs from both the original and its corresponding duplicate copy. The hysteresis of TH22 gates ensures the propagation of data through a stage if the register outputs of both the original and copied circuits are identical.
- Each register structure is modified as per [10] to accommodate two *Ki* signals instead of one, sourced from both the original and duplicate copies of the completion circuitry in the subsequent stage.
- Two additional *Illegal State Correction (ISC)* units are added to the output of the non-duplicated *CL* unit. The purpose and functionality of the *ISC* units are detailed later.
- The *CD* units are positioned at the output of the added TH22 gates, which produce *request-for-null* (*rfn, i.e., logic 0*) and *request-for-data* (*rfd, i.e., logic 1*) signals when all the TH22 outputs are DATA and NULL, respectively. For each stage, the *CD* unit drives the registration as well as the *ISC* units of both the original and duplicate circuits in the previous stages, as shown in Fig. 2.

As discussed earlier, the architecture is suitable for certain applications, such as signal and image processing, where a degree of approximation is acceptable and offers considerable performance benefits over exact/ accurate computing. Utilization of fast arithmetic units (e.g., advanced multipliers and adders) are common in such applications, where the data width exponentially increases the complexity of the circuit. It is straightforward to understand that a computational error in the least significant bits (LSBs) of adders/multipliers will exert a lesser influence on the overall output compared to an error in the most significant bits (MSBs). This comprehension facilitates the partitioning of the arithmetic logic into two units based on the preferred degree of approximation: a most significant unit (MSU) and a least significant unit (LSU). The objective is to guarantee the complete resilience of the MSU, maintaining an acceptable output quality, while permitting the LSU to remain unprotected to improve area and energy.

Herein, an NCL adder circuit is considered for explanation, where the *CL* unit is divided into two parts: $CL_{MSU}$ and $CL_{LSU}$ that generate the *(N-L)* most significant output bits and the *L* least significant bits, respectively. The signals and components that correspond to the original and duplicate copies are distinguished by the suffixes *(a)* and *(b)*, respectively. When an NCL pipeline contains maximum DATA tokens, two consecutive DATA stages are always separated by a single NULL stage [10]. Therefore, to evaluate the architecture during a SEU, we examine two potential scenarios: 1) when the affected stage is in a DATA state and 2) when it is in NULL state, prior to SEU, as elaborated next.

### A. Datapath Corruption: SEU Response under Scenario 1

Fig. 2 illustrates the response timeline under scenario 1, which occurs when SEU affects $stage_i$ components during a DATA state, while the subsequent stage is in a NULL state. A SEU can cause incorrect outputs in $stage_{i(a)}$ if it corrupts the datapath components of the stage., i.e., $CL_{MSUi(a)}$, $CL_{LSUi(a)}$, or $ISC_{i(a)}$. SEU can also affect the completion component, $CD_{i(a)}$, and produce an incorrect request signal. We demonstrate herein that the proposed architecture is capable of the following: fully recovering from incorrect output generation when SEU affects the $CL_{MSU}$ and *ISC* units; fully recovering from incorrect request signal generation without causing deadlock; and partially recovering when SEU affects the $CL_{LSU}$ unit by generating incorrect yet legal DATA, while keeping the computation error within an acceptable bound for the target application.

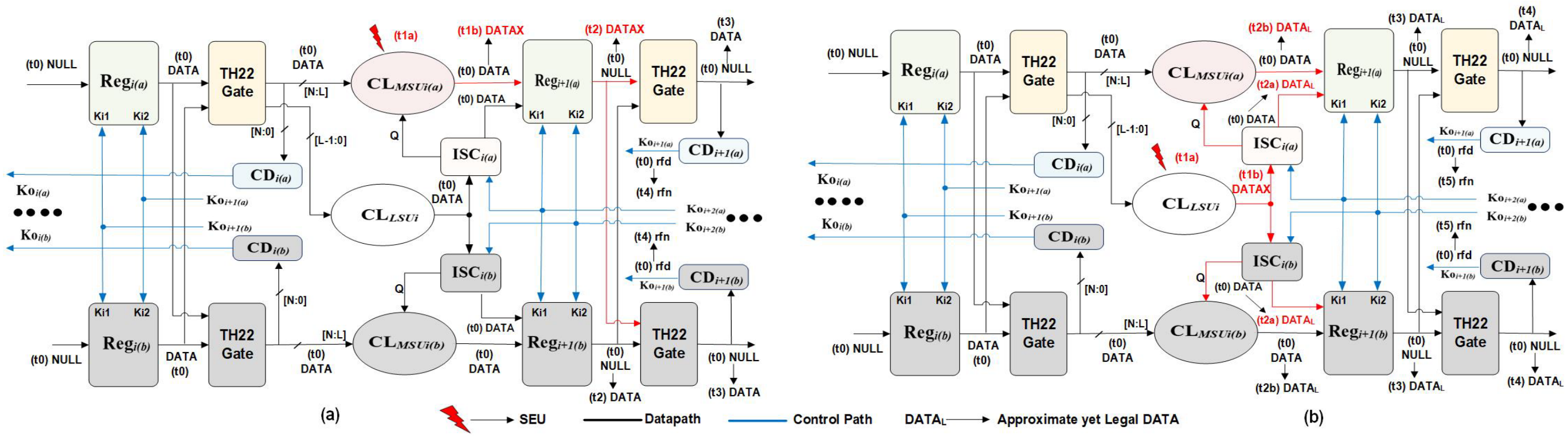

Fig. 2: Proposed SR-NCL framework; a) case I error scenario, and b) case II error scenario.

*Case I- SEU corrupting the $CL_{MSUi(a)}$ unit:* In Fig. 2a, assume a SEU at *t1a*, forcing undesired gate switching within the $CL_{MSUi(a)}$ unit to produce corrupted data at *t1b*. DATAX denotes corrupted data in which one rail of a signal is correctly asserted, while the other is either asserted or partially asserted. The DATAX output of $CL_{MSUi(a)}$ passes through $Reg_{i+1(a)}$ at *t2*, as both the CD units in subsequent stages are *rfd*. The matching register, $Reg_{i+1(b)}$, latches the correct DATA, as the duplicate copy remains unaffected. At *t3*, the TH22 gates in $stage_{i+1}$ will rectify the original copy's corrupted DATAX by using the correct DATA from the duplicate copy, and the circuit will fully recover. The DATAX in $stage_{i(a)}$ will eventually be flushed in the next NULL cycle.

*Case II- SEU corrupting the $CL_{LSUi}$ unit:* The $CL_{LSUi}$ unit is not duplicated and hence, any error in the unit propagates to both copies of the circuit. Therefore, in contrast to Case I, the duplicate circuit cannot be leveraged to rectify the corrupted DATA. Additionally, the corrupted output of both copies will be identical, which means that it will not be barred by the TH22 gates and will proceed to the next stage. To avoid the propagation of illegal DATA signals, two *ISC* units, one in each copy, are placed at the output of the $CL_{LSU}$ unit. The *ISC* unit functions as a buffer for valid DATA input, permitting it to pass at the output if the subsequent stage is *rfd*. If the subsequent stage is *rfn*, the *ISC* unit outputs NULL. If a corrupted DATAX is input to the unit, while the subsequent stage is *rfd*, the *ISC* output is forced to be a legal DATA (DATA0 or DATA1)[1], denoted as $DATA_L$. Consider a scenario in Fig. 2b, where a SEU affects the $CL_{LSUi}$ at *t1a*, and results in a corrupted DATAX at *t1b*. The *ISC* will convert that to $DATA_L$ in both copies at *t2*, which will eventually propagate to the next stages, as shown in *t3* and *t4*. Since the conversion of DATAX to $DATA_L$ is forceful and not based on re-computation, the output of the corrupted $CL_{LSUi}$ unit, in the worst case, will be incorrect yet legal. However, as this is an adder circuit, the replicated $CL_{MSU}$ units are connected to the $CL_{LSU}$ unit exclusively through an intermediary carry signal ($Q$), so the most significant output bits are mostly unaffected by the erroneous computation of the least significant unit. The *approximate* legal output, $DATA_L$, should be kept within an acceptable range for the target application, which can be adjusted by carefully tuning the width of the $CL_{LSU}$ unit.

*Case III- SEU corrupting the $ISC_{i(a)}$ unit:* To protect the correction unit, the *ISC* element is also doubled. The recovery process resembles Case I, wherein a corrupted $ISC_{i(a)}$ component may result in a DATAX state being latched by $Reg_{i+1(a)}$. However, the matching register, $Reg_{i+1(b)}$, will latch the correct DATA. This will enable the TH22 gates in $stage_{i+1}$ to filter out the corrupted DATA, similar to Case I.

### *B. Datapath Corruption: SEU Response under Scenario 2*

Scenario 2 transpires when SEUs impact $stage_i$ components during a NULL state, while the following stage, $stage_{i+1}$, is in a DATA state. Under this scenario, a SEU on the $CL_{MSUi(a)}$ unit can induce a bit-flip in one of the internal rails of a signal, leading to erroneous DATA0 or DATA1 outputs. (DATAX is infeasible in this context, as we are examining SEUs, and both rails of a NULL signal are not expected to be inverted simultaneously). However, since the duplicate copy remains intact and NULL, the incorrect DATA will not affect the subsequent stage. In addition, the TH22 gates in $stage_i$ will continue to hold the correct NULL, which will flush the incorrect DATA. Similarly, a SEU on the $CL_{LSU(i)}$ unit may also cause the unit to produce incorrect DATA0 or DATA1 values. However, the erroneous $CL_{LSU(i)}$ outputs are unable to update the *ISC* units in $stage_i$, which will remain NULL as the subsequent stage remains *rfn*. Finally, the $ISC_{i(a)}$ unit may momentarily output incorrect DATA0/DATA1 due to a SEU during the NULL phase. However, the $ISC_{i(a)}$ outputs will eventually be returned to NULL, as the succeeding stage remains *rfn*. Even if the following stage register captures the incorrect DATA produced by the corrupted $ISC_{i(a)}$, the TH22 gates will continue to block it in the subsequent stage, as the duplicate circuit holds the intact NULL.

---

[1] Our design of the ISC unit forces an illegal DATA to a DATA0 value.

### C. SEU Response under Control Path Corruption

In conventional NCL framework, the 4-phased handshaking protocol mandates that a register passes a NULL/DATA token to a specific stage only after the previous DATA/NULL token has been latched by the register in the succeeding stage. The protocol is maintained by the completion and registration units. A violation will occur if the *CD* unit within a stage prematurely requests the next NULL/DATA wavefront before the next stage register has completed latching the previous DATA/NULL. An error in the *CD* unit can induce such premature requests, which can eventually lead to a circuit deadlock. The proposed SR-NCL architecture can efficiently tackle such errors without causing deadlocks. Assume a scenario in Fig. 2, where at *t0*, the $stage_i$ C/L units are still computing based on an input DATA set and have only finished partial computation (i.e., some outputs have finished computing and are DATA, while the others are NULL). The next stage detection units, $CD_{i+1(a)}$ and $CD_{i+1(b)}$, are both *rfd*s as the complete DATA is not yet available. At $t_1$, a SEU corrupts the $CD_{i+1(a)}$ unit and prematurely changes the *rfd* signal to *rfn*. However, the erroneous request will not be sufficient for the $stage_i$ registers to pass the next NULL. This is because the intact $CD_{i+1(b)}$ still holds the correct *rfd* signal, and the modified register structures in the proposed design necessitate that both the request signals be *rfn* to pass a NULL. Eventually, at *t2*, the C/L units will finish computation and the complete DATA will be available at the $stage_{i+1}$ register outputs. The DATA will pass through the TH22 gates at *t3*, and at *t4* both $CD_{i+1(a)}$ and $CD_{i+1(b)}$ will be updated with the correct *rfn* signals. Similar recovery procedure will follow if $CD_{i+1(a)}$ produces a premature *rfd* signal before the complete NULL gets latched in $stage_{i+1}$.

## IV. Simulation Results and Discussion

Multiple SR-NCL carry look-ahead adder (CLA) circuits with different partitioning were designed and modeled in Verilog HDL. The functionality and recovery procedure of each circuit under various SEU conditions were validated through extensive simulation. One 8-bit SR-NCL CLA with 5-bit MSU and 3-bit LSU **(5|3)**, and two 16-bit SR-NCL CLAs, one with 11-bit MSU and 5-bit LSU **(11|5)** and one with 10-bit MSU and 6-bit LSU **(10|6)** were designed at the transistor level, utilizing high-performance 32-nm CMOS predictive technology model (PTM) [16]. A non-partitioned 8 and 16-bit DMR-NCL [10] CLA was also designed using the same model for performance comparison. The simulation results are summarized in Table I. The DMR-NCL CLAs and the proposed SR-NCL CLAs were evaluated and compared on three critical areas: transistor count, average power dissipation per operation, and the average DATA-to-DATA propagation delay ($T_{DD}$). For the 8-bit CLAs, results demonstrate that the proposed SR-NCL CLA achieved **~12%** and **~13%** reduction in transistor count and average power dissipation, respectively, in comparison to the DMR-NCL version, while maintaining a comparable latency. For the 16-bit CLAs, the proposed SR-NCL CLA with an 11-bit MSU and a 5-bit LSU produced reductions of approximately **8.8%** in transistor count and **10.3%** in average power consumption than the DMR-NCL version. The transistor count (**~11%**) and average power dissipation (**~13.1%**) were further reduced for the alternate design with a 10-bit MSU and 6-bit LSU.

Table I: Simulation Results

| Design | Transistor Count | Avg. Power Dissipation (µW) | Delay (ns) |
|---|---|---|---|
| 8-bit DMR- NCL CLA | 8,744 | 39.94 | 0.43 |
| **8-bit SR-NCL CLA (5\|3)** | **7,696** ↓ | **34.75** ↓ | **0. 43** |
| 16-bit DMR NCL CLA | 17,200 | 81.79 | 0. 69 |
| **16-bit SR-NCL CLA (11\|5)** | **15,683** | **73.33** | **0.74** |
| **16-bit SR-NCL CLA (10\|6)** | **15,309** ↓ | **71.10** ↓ | **0.76** ↑ |

The results illustrate that a larger width of the LSU can result in greater savings, albeit at the cost of a slightly increased latency. Designers must determine the optimal partition based on the target application's sensitivity to computing errors. To illustrate further, results of an image processing application have been presented in Fig 3. We target a 32-bit adder for this analysis, which is a common arithmetic component used during image reconstruction. Five different partition arrangements for the adder have been presented for this analysis: (**24|8**), (**22|10**), (**20|12**), (**19|13**), and (**18|14**), as depicted in Figs. 3b-f, respectively. In all arrangements, the carry signal generated by the LSU that propagates to the MSU has been considered corrupted to imitate a radiation induced error scenario. The quality of the images is assessed based on two parameters: Peak Signal-to-Noise Ratio (PSNR) and the Structural Similarity Index Measure (SSIM). A SSIM value of >0.85 is regarded as high-quality, 0.70< SSIM< 0.85 is regarded as acceptable, 0.30< SSIM<0.70 is regarded as low-quality, and SSIM<0.3 is regarded as poor quality [17]. Results suggest that the **(24|8)**, **(22|10),** and **(20|12)** adders generate high/acceptable quality images, despite the erroneous computation, whereas the **(19|13)** and **(18|14)** adders generate low/poor quality images. This demonstrates that a level of computational inaccuracy may still be acceptable for such applications, thereby validating the rationale behind our selective redundancy scheme.

## Acknowledgment

This work is supported by NSF; Grant No. CCF-2153373.

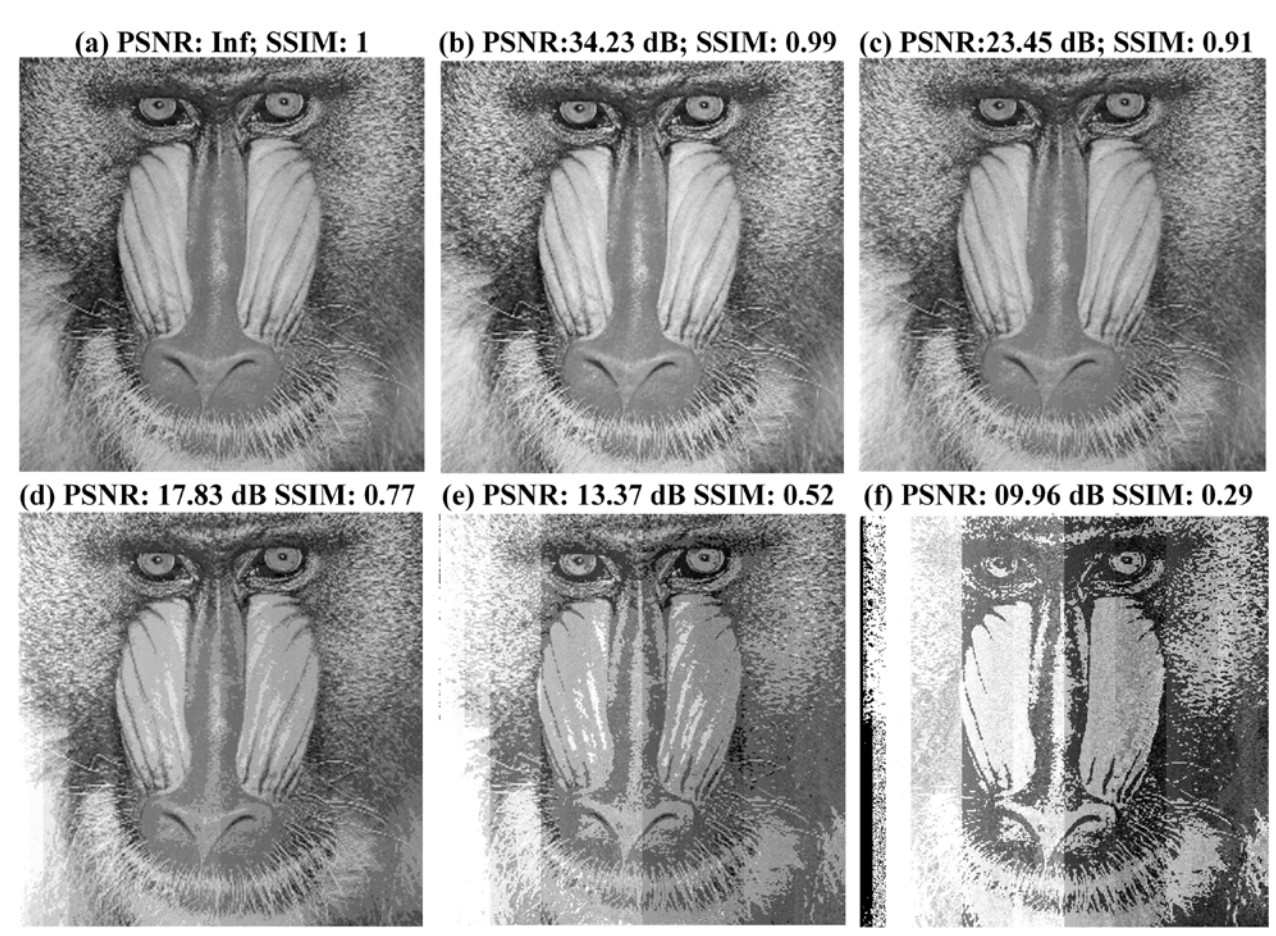

Fig 3: Image processing results using (a) accurate adder, and erroneous (b) 24|8, c) 22|10, (d) 20|12, (e) 19|13, and f) 18|14 adders.

## V. CONCLUSION

A novel scheme towards designing area-/energy efficient SEU tolerant NCL circuits has been proposed. The idea is based on selective redundancy where only a selective part of the circuit is duplicated. The non-duplicated portion remains unprotected against SEUs, whereas the duplicated portion provides full resilience. Simulation results indicate that deliberately reducing SEU resilience to an acceptable level for the target application can yield substantial reductions in area and energy utilization, while maintaining comparable latency and preserving the NCL protocols and deadlock-free operation.